\documentclass[superscriptaddress,twocolumn,amsmath,amssymb]{revtex4}
\usepackage{bm}
\usepackage{latexsym}
\usepackage[dvips]{color}
\usepackage{graphicx}
\usepackage{amsmath}
\usepackage{amssymb}
\usepackage{float}
\usepackage{dcolumn}
\begin{document}

\title{Single-enzyme kinetics with branched pathways:\\exact  theory and series expansion}

\author{Ashok Garai}
\affiliation{Centre for Condensed Matter Theory, Department of Physics, Indian Institute of Science, Bangalore 560012, India}
\author{Debashish Chowdhury}
\affiliation{Department of Physics, Indian Institute of Technology, Kanpur 208016, India}
\begin{abstract}
The progress of the successive rounds of catalytic conversion of 
substrates into product(s) by a single enzyme is characterized by 
the distribution of turnover times. Establishing the most general 
form of dependence of this distribution on the substrate concentration 
[S] is one of the fundamental challenges in single molecule enzymology. 
The distribution of the times of dwell of a molecular motor at the 
successive positions on its track is an analogous quantity. We 
derive approximate series expansions for the [ATP]-dependence of 
the first two moments of the dwell time distributions of motors that 
catalyze hydrolysis of ATP to draw input energy. Comparison between 
our results for motors with branched pathways and the corresponding 
expressions reported earlier for linear enzymatic pathways provides 
deep insight into the effects of the branches. Such insight is likely 
to help in discovering the most general form of [S]-dependence of 
these fundamental distributions.

\end{abstract}

\maketitle


Enzymes are proteins that catalyze chemical reactions.
The simplest, and most extensively studied, enzymatic reaction is 
\begin{equation}
E + S \mathop{\rightleftharpoons}^{\omega_{+1}}_{\omega_{-1}} ES \mathop{\rightarrow}^{\omega_{2}} E + P
\label{eq-MMkinetics}
\end{equation}
where $E$, $S$ and $P$ denote the enzyme, substrate and product, respectively.
The rate $V$ of the reaction in bulk was derived by Michaelis and Menten in a
celebrated classic \cite{MMoriginal} that marked its centenary last year 
\cite{FEBSJ,FEBSL}. 
A more general form of this rate $V$, derived by Briggs and Haldane 
about a decade later \cite{briggs25}, is now usually referred to as 
the Michaelis-Menten (MM) equation.

For a chemical reaction where a bulk of substrate is catalyzed by a 
{\it single molecule} of the corresponding enzyme 
\cite{claessen10,xie13,qian14}, 
rate is an ill-defined concept; instead the time taken by the enzyme 
for its successive turnovers is the prime quantity of interest. 
The statistical properties of the 
turnover is well characterized by the probability distribution of the 
turnover times (DTT). Deriving exact analytical expressions for this 
distribution, particularly its dependence on the substrate concentration 
$[S]$, has been one of the fundamental challenges in single-molecule 
enzymology. 

This work is motivated by a specific class of enzymes, called molecular 
motor \cite{chowdhury13a,chowdhury13b,kolomeisky13},  that catalyzes 
hydrolysis of ATP (and, therefore, referred to as ATPase) to draw input 
energy for their mechanical movement. 
The distribution of dwell times (DDT) of a motor characterizes its 
stochastic alternating pause-and-translocation along a filamentous track. 
The DDT of a molecular motor is the analog of the DTT of the enzymes 
\cite{moffitt14,chowdhury14}.  
Although we use the terminology of ATPase (and use the specific notation 
[ATP] instead of the generic symbol [S] for the substrate concentration), 
all the general conclusions drawn here remain equally valid for other 
enzymes with identical kinetic schemes.

The first two moments of DTT have been at the main focus of attention 
many recent works. The inverse of the mean turnover time 
$\langle t \rangle$ is the average rate $V$ of the reaction. 
Similarly the fluctuations in the turnover time is expressed by the 
{\it randomness parameter} $r$ \cite{moffitt14}. 
In wide varieties of situations the mean turnover time $\langle t \rangle$ 
has been found \cite{xie13,qian14,moffitt14} to obey the MM eqn.  
\begin{equation}
\frac{1}{V} = \langle t \rangle = \frac{1}{\omega_{2}} \biggl(1+\frac{K_M}{[ATP]}\biggr)
\label{eq-tavg}
\end{equation} 
where the Michaelis constant $K_M$ is given by 
\begin{equation}
K_M = \frac{\omega_{-1}+\omega_{2}}{\omega_{+1}}
\label{eq-Michaelis}
\end{equation}
So far as the dependence of $r$ on [ATP]  is concerned, an elegant 
expression reported in \cite{moffitt10a} (see also 
refs.\cite{jung10,chaudhury13}), is given by \cite{moffitt10b,moffitt14}
\begin{widetext}
\begin{equation}
r = \frac{\langle t^{2} \rangle - \langle t \rangle ^{2}}{\langle t \rangle^{2}}
=\left (1+\frac{[ATP]}{K_M}\right )^{-2}\left(\frac{1}{N_L}+2\frac{\alpha}{N_L N_S}\frac{[ATP]}{K_M}+\frac{1}{N_S}\left(\frac{[ATP]}{K_M}\right)^2 \right)
\label{eq-r}
\end{equation}
\end{widetext}
which involves three parameters $N_{L}, N_{S}$ and $\alpha$; from 
now onwards, we'll refer to eq.(\ref{eq-r}) as the MCB equation.

Although for an overwhelmingly large class of enzymatic reactions 
$\langle t \rangle$ and $r$ follow the general forms of eqs.(\ref{eq-tavg}) 
and (\ref{eq-r}), respectively \cite{xie13}, 
significant deviations from it have been observed in case of a few 
exceptional kinetic schemes (see ref.\cite{moffitt14} for a recent review). 
One of these exceptional schemes arose originally in ref.\cite{nosc,greulich07} 
in the context of a monomeric molecular motor kinesin KIF1A 
\cite{chowdhury13a,chowdhury13b,kolomeisky13}. 
For $\langle t \rangle$ and $r$ of this motor we derived \cite{garai11} 
the exact analytical expressions which, unfortunately, show no obvious 
similarity or connection with eqns. (\ref{eq-tavg}) and (\ref{eq-r}).

\begin{figure}[tb]
\begin{center}
\includegraphics[width=0.85\columnwidth]{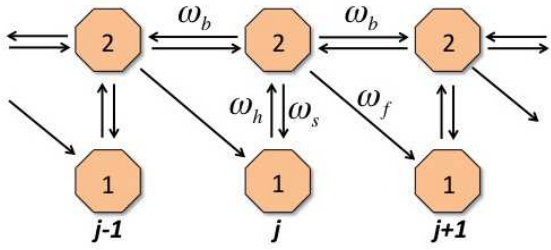}
\end{center}
\caption{(Color online) The kinetic scheme of our model is shown. The integers ...$j-1,j,j+1$... denote the {\it spatial} positions of the motor. At each spatial position the motor can exist in one of the two allowed ``internal'' states that are labelled by 1 and 2. The arrows and the associated symbols depict the allowed transitions and the corresponding rates. Note that in the limit $\omega_{b}=0$ this scheme reduces to the MM kinetics provided one identifies states $1$ and $2$ with 
the substrate-free and substrate-bound states of the enzyme and the rates $\omega_{h}, \omega_{s}, \omega_{f}$ with $\omega_{+1}, \omega_{-1}, \omega_{2}$, respectively, in equation (\ref{eq-MMkinetics}). }
\label{fig-kinscheme}
\end{figure}

The branched pathways in the kinetics of KIF1A \cite{nosc,garai11} 
(see fig.\ref{fig-kinscheme}) are suspected \cite{moffitt14} to cause 
deviations from the forms (\ref{eq-tavg}) and (\ref{eq-r}) for 
$\langle t \rangle$ and $r$. 
In this letter we show that the expressions (\ref{eq-tavg}) and 
(\ref{eq-r}) correspond to the first couple of terms in the series 
obtained by expanding the corresponding exact results of ref.\cite{garai11} 
in terms of the appropriate variables that we have now identified.
A comparison of the series expansions and exact expressions of 
$\langle t \rangle$ and $r$ for KIF1A with the corresponding expressions 
(\ref{eq-tavg}) and (\ref{eq-r}) highlight the contributions from the 
branched pathways in a systematic fashion. The new insight is likely 
to pave the way for eventual discovery of the most general expression 
for the [S]-dependence of $\langle t \rangle$ and $r$.

The full exact expression for $\langle t \rangle$ for the kinetic scheme shown in 
fig.\ref{fig-kinscheme} is given by \cite{garai11} 
\begin{widetext}
\begin{equation}
\langle t \rangle = \frac{(2\omega_b+\omega_f+\omega_h+\omega_s)\left \{ \omega_b\sqrt{(2\omega_b+\omega_f+\omega_s-\omega_h)^2+4\omega_h\omega_s}+\omega_f\omega_h-\omega_b(2\omega_b+\omega_f+\omega_s-\omega_h) \right\}}{\{(2\omega_b+\omega_f)\omega_h\}^2}
\label{eq-tavg-exact}
\end{equation}
\end{widetext}
which reduces to the simpler form (\ref{eq-tavg}) when $\omega_{b}=0$. 

For convenience, we introduce the symbols  
\begin{equation}
\gamma=\omega_s/(2\omega_b+\omega_f+\omega_s), 
\label{eq-gamma}
\end{equation} 
$y=(2\omega_b+\omega_f+\omega_s)/\omega_h$, and {\it assume} that 
\begin{equation}
(1-y)^2 \gg 4\gamma y  
\label{eq-assume1}
\end{equation}

The approximation (\ref{eq-assume1}) is valid if $\omega_{b}$ is 
sufficiently small and $\omega_{h}$ is much larger than both 
$\omega_{s}$ and $\omega_{f}$. Small $\omega_{b}$ helps in 
examining the connection with the limit $\omega_{b} = 0$.
Moreover, since 
\begin{equation}
\omega_{h} = \omega_{h}^{0} [ATP],  
\label{eq-secondorder}
\end{equation}
where $\omega_{h}^{0}$ is independent of ATP concentration, large 
$\omega_{h}$ corresponds to high concentration of ATP provided 
$\omega_{h}^{0}$ is not too small. 

Under the approximation (\ref{eq-assume1}), eq.(\ref{eq-tavg-exact}) 
reduces to   
\begin{widetext}
\begin{equation}
\langle t \rangle = \frac{1}{(2\omega_b+\omega_f)}\left \{1+\frac{2\gamma \omega_b+\omega_f}{2\omega_b+\omega_f}\frac{\kappa}{[ATP]} \right\}+\frac{2\omega_b}{(2\omega_b+\omega_f)^2}\biggl(\frac{\kappa}{[ATP]} \biggr)^{2} \sum^{M}_{m=0}\zeta_m\biggl(\frac{\kappa}{[ATP]} \biggr)^m,
\label{eq-tavg-appr}
\end{equation}
\end{widetext}
where the Michaelis-like constant $\kappa$ is given by 
\begin{equation}
\kappa = \frac{2\omega_b+\omega_f+\omega_s}{\omega^0_h}.
\label{eq-kappa1}
\end{equation}
and $M \to \infty$; the analytical expressions for the first five 
coefficients $\zeta_{m}$ ($m=0,1,..,4$) in (\ref{eq-tavg-appr}) 
are given in appendix A in terms of the rate constants $\omega$s. 
As $\omega_{b} \to 0$, $\kappa \to K_M$ and the approximate expression 
(\ref{eq-tavg-appr}) also reduces to the form (\ref{eq-tavg}).  

\begin{figure}[H]
\includegraphics[width=0.85\columnwidth]{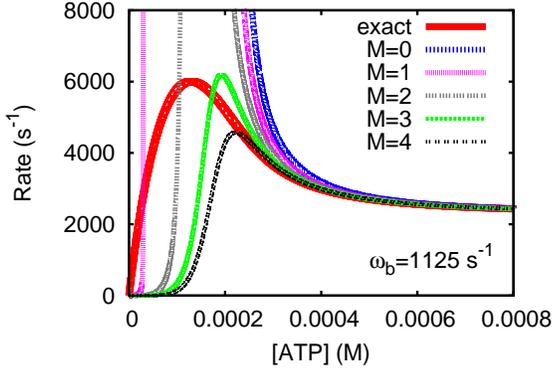}\\
\caption{(Color online) Average ATPase rate is plotted as a function of ATP 
concentration using inverse of the Eq.(\ref{eq-tavg-exact}) (red), and Eq.(\ref{eq-tavg-appr}) 
(dotted lines for $M=0 \text{(Blue)}, 1 \text{(Magenta)}, 2 \text{(Grey)}, 3 \text{(Green)}, 4 \text{(Black)}$) for $\omega_b=1125 s^{-1}$. 
Other parameters are as follows: $\omega_f=55 s^{-1}$ and $\omega_s=145 s^{-1}$.
}
\label{fig-series}
\end{figure}

In fig.\ref{fig-series} we have plotted the predictions of the series 
in (\ref{eq-tavg-appr}) for $M=0,1,2,3,4$ and compared these with the 
corresponding prediction of the exact expression (\ref{eq-tavg-exact}). 
The higher is the value of $M$, the better is the agreement with the 
exact result.

\begin{figure}[H]
\includegraphics[width=0.85\columnwidth]{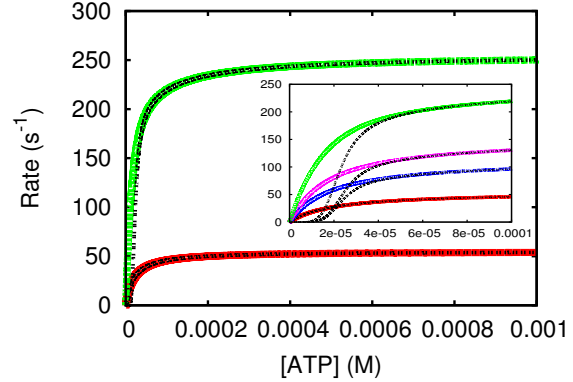}\\
\caption{(Color online) Average ATPase rates obtained from the 
exact expression (\ref{eq-tavg-exact}) are plotted against [ATP] 
by continuous curves
for $\omega_b=0$ (red), $\omega_b=30 s^{-1}$ (blue), $\omega_b=50 s^{-1}$ 
(magenta) and $\omega_b=100 s^{-1}$ (green)).
Black dotted lines for the respective $\omega_b$ values are obtained from 
the approximated expression (\ref{eq-tavg-appr}). Other 
parameters used for this plot are $\omega_s=145 s^{-1}, \omega_f=55 s^{-1}$.  
}
\label{fig-avt}
\end{figure}

For a graphical test of the range of validity of the approximate 
expression (\ref{eq-tavg-appr}) we plot the inverse of $\langle t \rangle$ 
(i.e., the average rate) in Fig.\ref{fig-avt} as a function of ATP 
concentration for a few different values of $\omega_{b}$.  
Even with $M$ as low as $4$, the predictions of (\ref{eq-tavg-appr}) 
are practically indistinguishable from those of (\ref{eq-tavg-exact}), except 
at very low [ATP] (see the inset of Fig.\ref{fig-avt}).

The exact expression for $r$ in this model is \cite{garai11}  
\begin{widetext}
\begin{equation}
r = \frac{2(2\omega_b+\omega_f)\omega_h\biggl[ (2\omega_b+\omega_f)\omega_h-(2\omega_b+\omega_f+\omega_h+\omega_s)^2 \biggr]}{(2\omega_b+\omega_f+\omega_h+\omega_s)^2 \biggl[\omega_b(2\omega_b+\omega_f+\omega_s-\omega_h)-\omega_f\omega_h-\omega_b\sqrt{(2\omega_b+\omega_f+\omega_s-\omega_h)^2+4\omega_h\omega_s} \biggr]}-1
\label{eq-r-exact}
\end{equation}
\end{widetext}

\begin{figure}[H]
\includegraphics[width=0.85\columnwidth]{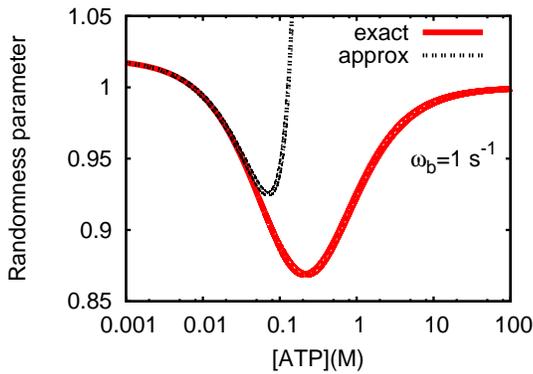}\\
\caption{(Color online) Randomness parameter evaluated from the 
approximate expression (\ref{eq-r-appr5}) for $\omega_{b}=1$ s$^{-1}$ 
is compared with the corresponding prediction of the exact expression 
(\ref{eq-r-exact}) by plotting both against [ATP]. The other parameters 
used for this figure are $\omega_s=145$s$^{-1}$, $\omega_f=55$s$^{-1}$, 
and $\omega^0_h=1000$s$^{-1}$ per mole.  
}
\label{fig-exactrapprox}
\end{figure}

In all the plots made above we used $\omega_{h}^{0} = 10^7$s$^{-1}$ 
per mole. But, for approximating (\ref{eq-r-exact}) by an infinite 
series that correlates directly with eq.(\ref{eq-r}), we now also 
{\it assume} that $\omega_{h}^{0}$ is sufficiently small so that 
\begin{equation}
[ATP]/\kappa \ll 1, ~~~{\rm in ~spite ~of ~high}~~  [ATP]. 
\label{eq-assume2}
\end{equation}
Under the approximations (\ref{eq-assume1}) eq.(\ref{eq-r-exact}) reduces to 
\begin{equation}
r=\biggl(1+\frac{[ATP]}{\kappa}\biggr)^{-2} \sum^{\infty}_{n=0} \xi_n\biggl(\frac{[ATP]}{\kappa} \biggr)^n;
\label{eq-r-appr5}
\end{equation}
the expressions for the first six coefficients $\xi_{n} (n=0,1,..,5)$ 
are given in the appendix B. The series in eq.(\ref{eq-r-appr5}) 
converges rapidly because of the condition (\ref{eq-assume2}) and it 
reduces to eq.(\ref{eq-r}) for $\omega_{b}=0$. 
Over physiologically relevant range of concentration of ATP, 
which hardly ever exceeds few mM, the expression (\ref{eq-r-appr5}) 
can yield highly accurate estimate of $r$, as demonstrated by the 
graphical comparison with the corresponding  exact result 
(\ref{eq-r-exact}) in fig.\ref{fig-exactrapprox}). 

In almost all branches of physical sciences series expansion has been 
an extremely powerful tool for systematic approximation of important 
quantities. In the same spirit in this letter we have introduced 
series expansion method for analyzing stochastic kinetics in 
single-molecule enzymology and single-motor biophysics. We have also 
established the utility of these {\it generalized} MM and MCB equations 
\cite{wu12} by comparing these with the corresponding exact results. 
The series expansion approach is expected to be very useful if exact 
analytical treatment is impossible because of the complexity of the 
network of pathways in the mechano-chemical kinetics of the system.


\begin{appendix}

\section{Coefficients of the series for $\langle t \rangle$} 

\begin{equation}
\zeta_0 = - (1-\gamma)^2,
\label{eq-tavg-zeta0}
\end{equation}

\begin{equation}
\zeta_1 = 2\gamma(1-\gamma)^2,
\label{eq-tavg-zeta1}
\end{equation}

\begin{equation}
\zeta_2 = \gamma(2-5\gamma)(1-\gamma)^2,
\label{eq-tavg-zeta2}
\end{equation}

\begin{equation}
\zeta_3=2\gamma(1-6\gamma+7\gamma^2)(1-\gamma)^2,
\label{eq-tavg-zeta3}
\end{equation}

\begin{equation}
\zeta_4=\gamma\biggl(2+5\gamma\biggl[\gamma\left\{20+7\gamma(4\gamma-5) \right\} -5 \biggr] \biggr)
\label{eq-tavg-zeta4}
\end{equation}
where $\gamma$ is given by eq.(\ref{eq-gamma}).

\section{Coefficients of the series for $r$} 

\begin{equation}
\xi_0 = \frac{4\omega_b+\omega_f+\omega_s}{\omega_f+\omega_s}
\label{eq-r-xi0-w}
\end{equation}

\begin{equation}
\xi_1 = \frac{2\biggl(2\omega_b\omega_f + \omega_s(\omega_f+\omega_s) \biggr)}{(\omega_f+\omega_s)^2}
\label{eq-r-xi1-w}
\end{equation}

\begin{widetext}
\begin{equation}
\xi_2 = \frac{16\omega^3_b\omega^2_f+\omega_f(\omega_f+\omega_s)^4+2\omega_b(\omega_f+\omega_s)(4\omega_f+5\omega_s)(\omega^2_f+\omega^2_s)+4\omega^2_b(5\omega^3_f+5\omega^2_f\omega_s+3\omega_f\omega^2_s+\omega^3_s)}{(2\omega_b+\omega_f)(\omega_f+\omega_s)^3(2\omega_b+\omega_f+\omega_s)}
\label{eq-r-xi2-w}
\end{equation}

\begin{eqnarray}
\xi_3 &=&\frac{4\omega_b\omega_s}{(2\omega_b+\omega_f)(\omega_f+\omega_s)^4(2\omega_b+\omega_f+\omega_s)^2} \biggl[ -4\omega_b^2\omega_f(3\omega_f+2\omega_s)^2 - 2\omega_b(\omega_f+\omega_s)(3\omega_f+2\omega_s)(3\omega^2_f-4\omega^2_s) \nonumber \\
&-& 3(\omega_f+\omega_s)^3(\omega^2_f-\omega_f\omega_s-3\omega^2_s) - 8\omega^3_b(3\omega^2_f+3\omega_f\omega_s+\omega^2_s) \biggr]
\label{eq-r-xi3-w}
\end{eqnarray}

\begin{eqnarray}
\xi_4 &=& \frac{4\omega_b\omega_s}{(2\omega_b+\omega_f)^2(\omega_f+\omega_s)^5(2\omega_b+\omega_f+\omega_s)^3}\biggl[ -3\omega^3_f(2\omega_b+\omega_f)^5 - 3\omega^2_f(2\omega_b+\omega_f)^5\omega_s \nonumber \\
&+& \omega_f(2\omega_b+\omega_f)^3(16\omega_b\omega_f-4\omega^2_b+51\omega^2_f)w_s^2 + \omega_f(2\omega_b+\omega_f)^2(40\omega^2_b+246\omega_b\omega_f+195\omega^2_f)\omega^3_s \nonumber \\
&+& (2\omega_b+\omega_f)(24\omega^3_b+348\omega^2_b\omega_f+716\omega_b\omega^2_f+315\omega^3_f)\omega^4_s + (160\omega^3_b+840\omega^2_b\omega_f+922\omega_b\omega^2_f+267\omega^3_f)\omega^5_s \nonumber \\
&+& (160\omega^2_b+330\omega_b\omega_f+117\omega^2_f)\omega^6_s+(50\omega_b+21\omega_f)\omega^7_s\biggr]
\label{eq-r-xi4-w}
\end{eqnarray}

\begin{eqnarray}
\xi_5 &=& \frac{4\omega_b\omega_s}{(2\omega_b+\omega_f)^2(\omega_f+\omega_s)^6(2\omega_b+\omega_f+\omega_s)^4}\biggl[-3\omega^4_f(2\omega_b+\omega_f)^6 + 3\omega^3_f(\omega_f-2\omega_b)(2\omega_b+\omega_f)^5\omega_s \nonumber \\
&+& 2\omega^2_f(2\omega_b+\omega_f)^4(31\omega_b\omega_f-2\omega^2_b+71\omega^2_f)\omega^2_s + 2\omega^2_f(2\omega_b+\omega_f)^3(76\omega^2_b+404\omega_b\omega_f+329\omega^2_f)\omega^3_s  \nonumber \\
&+& 12\omega_f(2\omega_b+\omega_f)^2(12\omega^3_b+135\omega^2_b\omega_f+271\omega_b\omega^2_f+126\omega^3_f)\omega^4_s +2(2\omega_b+\omega_f)(24\omega^4_b+732\omega^3_b\omega_f+3034\omega^2_b\omega^2_f \nonumber \\
&+& 3365 \omega_b\omega^3_f+1036\omega^4_f)\omega^5_s+2(256\omega^4_b+2580\omega^3_b\omega_f+5644\omega^2_b\omega^2_f+4027\omega_b\omega^3_f+889\omega^4_f)\omega^6_s+6(140\omega^3_b \nonumber \\
&+& 626\omega^2_b\omega_f+628\omega_b\omega^2_f+157\omega^3_f)\omega^7_s+(556\omega^2_b+1036\omega_b\omega_f+283\omega^2_f)\omega^8_s+(130\omega_b+37\omega_f)\omega^9_s\biggr]
\label{eq-r-xi5-w}
\end{eqnarray} 
\end{widetext}

\end{appendix}

\section*{Acknowledgements}

DC thanks Prabal Maiti and the Physics department of Indian Institute of Science
for hospitality in Bangalore during preparation of this manuscript . This work has been supported by a J.C. Bose National Fellowship (DC) and a research grant from DAE (AG).



\end{document}